\documentclass[aps,twocolumn,superscriptaddress]{revtex4-1}

\usepackage{amsmath}
\usepackage{amsfonts}
\usepackage{longtable}
\usepackage[dvipdfmx]{graphicx}
\usepackage{isomath}

\begin{document}

\title{
A coherent Ising machine for maximum cut problems : Performance evaluation against semidefinite programming relaxation and simulated annealing}
\author{Yoshitaka Haribara}
\email[]{haribara@nii.ac.jp}
\affiliation{Department of Mathematical Informatics, The University of Tokyo, Hongo 7-3-1, Bunkyo-ku, Tokyo 113-8656, Japan}
\affiliation{National Institute of Informatics, Hitotsubashi 2-1-2, Chiyoda-ku, Tokyo 101-8403, Japan}

\author{Shoko Utsunomiya}
\email[]{shoko@nii.ac.jp}
\affiliation{National Institute of Informatics, Hitotsubashi 2-1-2, Chiyoda-ku, Tokyo 101-8403, Japan}

\author{Ken-ichi Kawarabayashi}
\email[]{k\_keniti@nii.ac.jp}
\affiliation{National Institute of Informatics, Hitotsubashi 2-1-2, Chiyoda-ku, Tokyo 101-8403, Japan}

\author{Yoshihisa Yamamoto}
\email[]{yyamamoto@stanford.edu}
\affiliation{E. L. Ginzton Laboratory, Stanford University, Stanford, CA94305, USA}
\affiliation{ImPACT program, The Japan Science and Technology Agency, Gobancho 7, Chiyoda-ku, Tokyo 102-0076, Japan}

\date{\today}

\begin{abstract}
Combinatorial optimization problems are computationally hard in general, but they are ubiquitous in our modern life. A coherent Ising machine (CIM) based on a multiple-pulse degenerate optical parametric oscillator (DOPO) is an alternative approach to solve these problems by a specialized physical computing system. To evaluate its potential performance, computational experiments are performed on maximum cut (MAX-CUT) problems against traditional algorithms such as semidefinite programming relaxation of Goemans-Williamson and simulated annealing by Kirkpatrick, et al. 
The numerical results empirically suggest that the almost constant computation time is required to obtain the reasonably accurate solutions of MAX-CUT problems on a CIM with the number of vertices up to $2 \times 10^4$ and the number of edges up to $10^8$.
\end{abstract}

\pacs{}

\maketitle

\section{\label{sec:intro}Introduction}
Combinatorial optimization appears in many important fields such as computer science \cite{karp1972complexity, mezard1987spin}, drug discovery and life-science \cite{kitchen2004docking}, and information processing technology \cite{nishimori2001statistical}. 
One of the example of such problems is an Ising problem to minimize the Ising Hamiltonian, which is a function of a spin configuration $\sigma = (\sigma_i)$ defined as
\begin{eqnarray}\label{eq:ising}
\mathcal{H}(\sigma) = - \sum_{i < j}J_{ij}\sigma_i\sigma_j\quad (1 \leq i,j \leq N),
\end{eqnarray}
where each spin takes binary values $ \sigma_i \in \{\pm 1\}$, a real number symmetric matrix $J_{ij}$ denotes a coupling constant, and $N$ is the total number of spins. 
Despite its simple statement, it belongs to the non-deterministic polynomial-time (NP)-hard class to find the ground state of the Ising model on the three-dimensional lattice \cite{barahona1982computational}. 

Similarly, a maximum cut (MAX-CUT) problem in the graph theory is to find the size of the largest cut in a given undirected graph. Here, a cut is a partition of the vertices $V$ into two disjoint subsets $\{S_1, S_2\}$ and the size of the cut is the total weight of edges $w_{ij}$ with one vertex $i$ in $S_1$ and the other $j$ in $S_2$. The size of the cut can be counted by assigning the binary spin values to express which subset the vertex $i$ belongs to $\sigma_i \in \{\pm 1\}$ \cite{johnson1979computers}:
\begin{eqnarray}\label{eq:cut}
C(\sigma) &=& \sum_{i \in S_1, j \in S_2}w_{ij} = \sum_{i<j}w_{ij} \frac{(1 - \sigma_i \sigma_j)}{2} \nonumber\\
&=& \frac{1}{2}\sum_{i<j}w_{ij} - \frac{1}{2}\mathcal{H}(\sigma),
\end{eqnarray}
where $\mathcal{H}$ is an Ising Hamiltonian defined in Eq.~(\ref{eq:ising}) with $J_{ij} = -w_{ij}$. It indicates that the MAX-CUT problem is equivalent to the Ising problem except for the constant factor. 

The MAX-CUT problem belongs to the NP-hard class in general, even though there are graph topologies which can be solved in polynomial time \cite{barahona1982computational, orlova1972finding, hadlock1975finding, grotschel1981weakly, grotschel1984polynomial, galluccio2001optimization}. Many attempts have been made to approximately solve NP-hard MAX-CUT problems, but the probabilistically checkable proof (PCP) theorem states that no polynomial time algorithms can approximate MAX-CUT problems better than $0.94118$ \cite{arora1998proof, haastad2001some}.
Currently, the approximation ratio of $0.87856$ achieved by the Goemans-Williamson algorithm (GW) based on semidefinite programming (SDP) is the best value for performance guarantee \cite{goemans1995improved}. This algorithm is a well-established benchmark to evaluate any new algorithms or computing methods.

Besides, there exist several heuristic algorithms to tackle these NP-hard MAX-CUT problems. The simulated annealing (SA) was designed by mimicking the thermal annealing procedure in metallurgy \cite{kirkpatrick1983optimization}. A quantum annealing technique was also formulated and was shown to have competitive performance against SA \cite{kadowaki1998quantum, santoro2002theory, farhi2001quantum, van2001powerful, aharonov2008adiabatic}. 
Independently, novel algorithms which are superior either in its speed or its accuracy are proposed \cite{sahni1976p, kahruman2007greedy, benlic2013breakout}.

We recently proposed a novel computing system to implement the NP-hard Ising problems using the criticality of laser \cite{utsunomiya2011mapping, takata2012transient, takata2014data, utsunomiya2015binary} and degenerate optical parametric oscillator (DOPO) phase transition \cite{wang2013coherent, marandi2014network}. The architecture of this machine is motivated by the principle of laser and DOPO in which the mode with the minimum loss rate is most likely to be excited first. The energy of the Ising Hamiltonian can be mapped onto the total loss rate of the laser or DOPO network. The selected oscillation mode in the laser or DOPO network corresponds to the ground state of a given Ising Hamiltonian, while the gain accessible to all other possible modes is depleted due to the cross-gain saturation. This means that a mode with the lowest loss rate reaches a threshold condition first and clumps the gain at its loss rate, so that all the other modes with higher loss rates stay at sub-threshold conditions. 
Moreover, the DOPO is in the linear superposition of $0$-phase state and $\pi$-phase state at its oscillation threshold \cite{drummond1980}. The coupled DOPOs form quantum entanglement in spite of their inherent dissipative natures \cite{takata2015quantum}, so that some form of quantum parallel search could be embedded in a DOPO network.

In this article, the validity of the CIM for MAX-CUT problems is tested against the representative approximation algorithms. The DOPO signal pulse amplitudes in CIM, which are interpreted as the solution, are described by the c-number stochastic differential equations (CSDE) as presented in Section \ref{sec:MF}. Then we conduct numerical simulations for MAX-CUT problems in Section \ref{sec:maxcut} with the number of vertices up to $N = 20000$. 
It is, of course, difficult to compare the performance of the proposed system as a MAX-CUT solver with the representative approximation algorithms which can be run on current digital computers mainly because the unit of ``clock'' cannot be uniquely defined. Thus we defined the feasible system clock which dominates the computing process in CIM as mentioned later. Moreover, here we evaluated the computational ability under either time or accuracy was fixed, while a preliminary benchmark study done in the previous paper is focused on the performance after physical convergence \cite{marandi2014network}. 

\section{\label{sec:MF}Multiple-pulse DOPO with mutual coupling}
\subsection{\label{sec:spec}A proposed machine}
\begin{figure*}
\includegraphics[width=4.in]{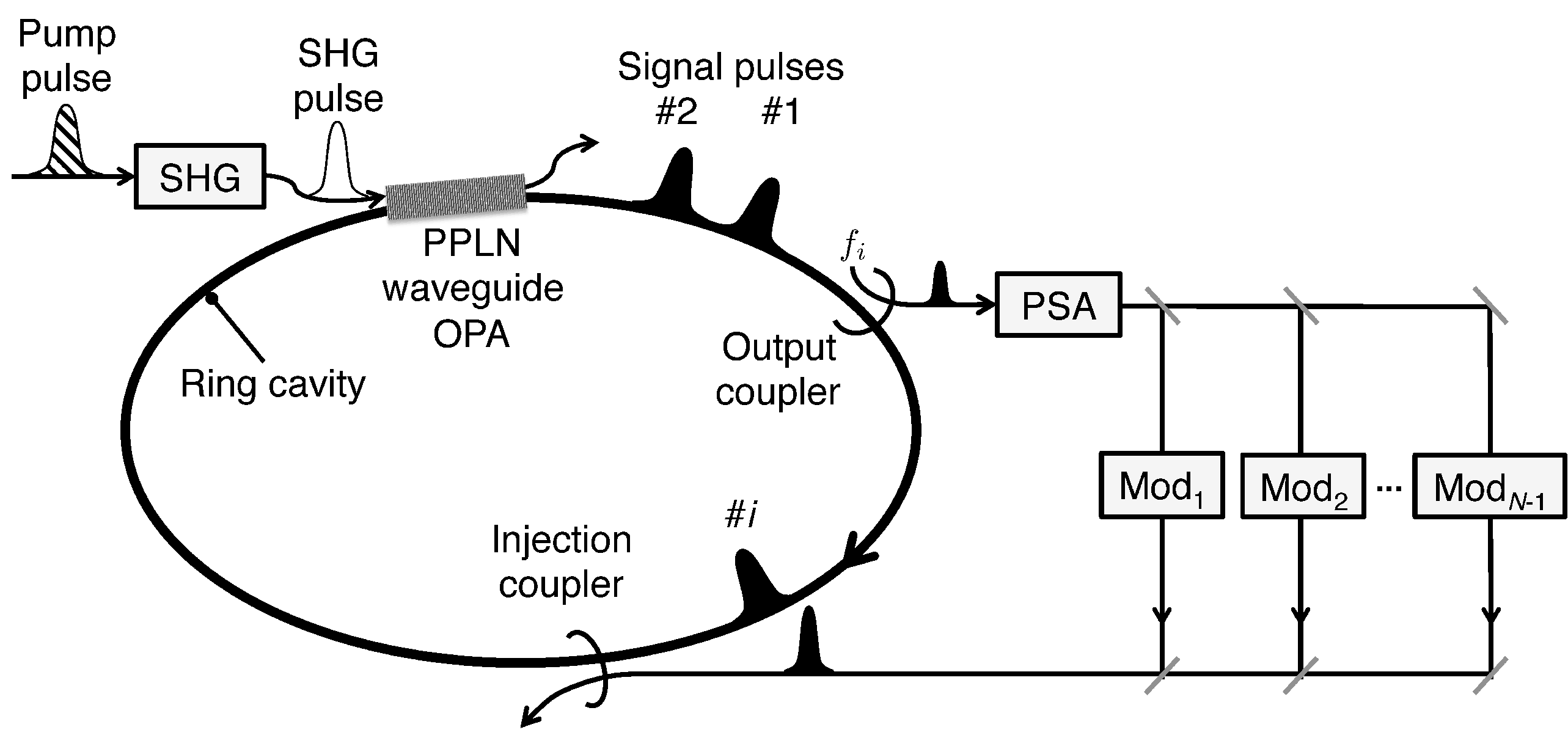}
\caption{\label{fig:CIM}A coherent Ising machine based on the time-division multiplexed DOPO with mutual coupling implemented by optical delay lines. A part of each pulse is picked off from the main cavity by the output coupler followed by an optical phase sensitive amplifier (PSA) which amplifies the in-phase amplitude $\tilde{c}_i$ of each DOPO pulse. The feedback pulses, which are produced by combining the outputs from $N-1$ intensity and phase modulators, are injected back to the main cavity by the injection coupler.}
\end{figure*}

A standard CIM based on multiple-pulse DOPO with all-optical mutual coupling circuits is shown in Fig. \ref{fig:CIM}. The system starts with a pulsed master laser at a wavelength of $1.56\ \mu \mathrm{m}$. A second harmonic generation (SHG) crystal produces the pulse trains at a wavelength of $0.78\ \mu \mathrm{m}$ which in turn generate multiple DOPO pulses at a wavelength of $1.56\ \mu \mathrm{m}$ inside a fiber ring resonator. If the round trip time of the fiber ring resonator is properly adjusted to $N$ times the pump pulse interval, we can simultaneously generate $N$ independent DOPO pulses inside the resonator. Each of these pulses is either in $0$-phase state or $\pi$-phase state at well above the oscillation threshold and represents an Ising spin of up or down.

In order to implement an Ising coupling $J_{ij}$ in Eq. (\ref{eq:ising}), a part of each DOPO pulse in the fiber ring resonator is picked-off and fed into an optical phase sensitive amplifier (PSA), followed by optical delay lines with intensity and phase modulators. Using such $N-1$ optical delay lines, (arbitrary) $i$-th pulse can be coupled to (arbitrary) $j$-th pulse with a coupling coefficient $J_{ij}$.
Such an all-optical coupling scheme has been demonstrated for $N = 4$ and $N = 16$ CIMs \cite{marandi2014network, kenta2015}.

In Sections \ref{sec:SDP} and \ref{sec:SA}, we assume a CIM with a fiber length of $2\ \mathrm{km}$ (or cavity round trip time of 10 $\mu$s) and pulse spacing of 10 cm (or pulse repetition frequency of $2\ \mathrm{GHz}$), thus $2\times10^{4}$ independent DOPO pulses can be prepared for computation. The system clock frequency for the CIM should be defined by the cavity circulation frequency (inverse of cavity round trip time). One clock cycle (round trip) includes every elements of computation, such as parametric amplification, out-coupling port, and coherent feedback. Thus the clock frequency of the CIM assumed for the present benchmark study is 100 kHz since the round trip time of 2 km fiber ring is $10\ \mu\mathrm{s}$. We fixed this system clock frequency, just like any digital computer has a fixed clock frequency and chose the appropriate pulse interval to pack the desired number of pulses in the $2 \mathrm{km}$ fiber. 

\subsection{c-number stochastic differential equations for multiple-pulse DOPO with mutual coupling} 
The in-phase and quadrature-phase amplitudes of a single isolated DOPO pulse obey the following c-number stochastic differential equations (CSDE) \cite{kinsler1991quantum}: 
\begin{eqnarray}
dc&=& (-1+p-c^2-s^2)c\, dt+\frac{1}{A_\mathrm{s}}\sqrt{c^2\!+\!s^2\!+\!\frac{1}{2}}dW_1\label{eq:1 CSDE} \\
ds&=& (-1-p-c^2-s^2)s\, dt+\frac{1}{A_\mathrm{s}}\sqrt{c^2\!+\!s^2\!+\!\frac{1}{2}}dW_2.\label{eq:2 CSDE}
\end{eqnarray}
The above CSDE are derived by expanding the DOPO field density operator with the truncated Wigner distribution functions. 
An alternative approach is to use two coherent states in the generalized (off-diagonal) $P(\alpha_\mathrm{s}, \beta_\mathrm{s})$-representation for the field density matrix \cite{drumond1981non}. The two approaches by the truncated Wigner function and the generalised P-representation are equivalent for highly dissipative systems such as ours. The pump field is adiabatically eliminated in (\ref{eq:1 CSDE}) and (\ref{eq:2 CSDE}) by assuming that the pump photon decay rate $\gamma_\mathrm{p}$ is much larger than the signal photon decay rate $\gamma_\mathrm{s}$. The term $A_\mathrm{s}=(\gamma_\mathrm{s} \gamma_\mathrm{p} /2 \kappa^2)^{1/2}$ is the DOPO field amplitude at a normalized pump rate $p = F_\mathrm{p}/F_{\mathrm{th}}=2$, and $\kappa$ is the second order nonlinear coefficient associated with the degenerate optical parametric amplification. The variable $t=\gamma_\mathrm{s}\tau / 2$ is a normalized time, while $\tau$ is a real time in seconds. The term $F_\mathrm{p}$ is the pump field amplitude and $F_{\mathrm{th}}=\gamma_\mathrm{s} \gamma_\mathrm{p}/4\kappa$ is the threshold pump field amplitude. Finally, $dW_1$ and $dW_2$ are two independent Gaussian noise processes that represent the incident vacuum fluctuations from the open port of the output coupler and the pump field fluctuation for in-phase and quadrature-phase components, respectively. The vacuum fluctuation of the signal channel contributes to the 1/2 term and the quantum noise of the pump field contributes to $c^2+s^2$ in the square-root bracket in (\ref{eq:1 CSDE}) and (\ref{eq:2 CSDE}).

When the $i$-th signal pulse is incident upon the output coupler, 
the output-coupled field and remaining field inside a cavity are written as
\begin{eqnarray}
c_{i, \mathrm{out}}&=&\sqrt{T}c_i-\sqrt{1-T}\frac{f_i}{A_\mathrm{s}}\label{eq:3 coupler 1}\\
c_{i, \mathrm{cavity}}&=&\sqrt{1-T}c_i+\sqrt{T}\frac{f_i}{A_\mathrm{s}}\label{eq:4 coupler 1},
\end{eqnarray}
where $T$ is the power transmission coefficient of the output coupler and $f_i$ is the incident vacuum fluctuation from the open port of the coupler. The out-coupled field and the signal amplitude after PSA can be
\begin{eqnarray}
\tilde{c}_{i} \equiv \frac{c_{i, \mathrm{out}}}{\sqrt{T}}  = c_i-\sqrt{\frac{1-T}{T}}\frac{f_i}{A_\mathrm{s}}.\label{eq:5 coupler 1_}
\end{eqnarray}
From these out-coupled pulse stream, the intensity and phase modulators placed in the $N-1$ delay lines produce the mutual coupling pulse $\sum_j \xi_{ij}\tilde{c}_j$, which is actually added to the $i$-th signal pulse by an injection coupler. Here, $\xi_{ij}$ is the effective coupling coefficient from the $j$-th pulse to the $i$-th pulse, determined by the transmission coefficient $\sqrt{T'}$ of the injection coupler. In the highly dissipative limit of a mutual coupling circuit, such as in our scheme, we can use the CSDE supplemented with the noisy coupling term. Since the transmission coefficient $\sqrt{T'}$ of the injection coupler should be much smaller than one, we do not need to consider any additional noise in the injected feedback pulse. The CSDE (\ref{eq:1 CSDE}) can be now rewritten to include the mutual coupling terms
\begin{eqnarray}
dc_{i}=[(-1+p-c_i^2-s_i^2)c_i + \sum_{j}\xi_{ij}\tilde{c}_j]\, dt \nonumber\\
+\frac{1}{A_\mathrm{s}}\sqrt{c_i^2+s_i^2+\frac{1}{2}}dW_i.\label{eq:6 coupler 1,2}
\end{eqnarray}
The summation in Eq. (\ref{eq:6 coupler 1,2}) represents the quantum measurement-feedback term including the measurement error given by Eq. (\ref{eq:5 coupler 1_}). The vacuum fluctuation coupled to the $i$-th pulse in the output coupler is already taken into account in the last term of right-hand side of Eq. (\ref{eq:6 coupler 1,2}) together with the pump noise. 
We conducted the numerical simulation of the coupled CSDE (\ref{eq:6 coupler 1,2}) to evaluate the performance of the CIM.

\section{\label{sec:maxcut}Benchmark studies on MAX-CUT problems}
\subsection{MAX-CUT problems on cubic graphs} 
The MAX-CUT problem on cubic graphs, in which each vertex has exactly three edges, is called MAX-CUT-3 problem and also belongs to NP-hard class \cite{halperin2002max}. The smallest simple MAX-CUT-3 problem is defined on the complete graph $K_4$ with four vertices and six edges with identical weight $J_{ij} = -1$, where anti-ferromagnetic couplings have frustration so that the ground states are highly degenerate. The solution to this problem are the set of two-by-two cuts, which contains six degenerate ground states of the Ising Hamiltonian, i.e., $\{\left|\uparrow \uparrow \downarrow \downarrow \right\rangle, \left|\uparrow \downarrow \uparrow \downarrow \right\rangle, \left|\uparrow \downarrow \downarrow \uparrow \right\rangle, \left|\downarrow \uparrow \uparrow \downarrow \right\rangle, \left|\downarrow \uparrow \downarrow \uparrow \right\rangle, \left|\downarrow \downarrow \uparrow \uparrow \right\rangle\}$.

\begin{figure}[htbp]
\centering
\includegraphics[width=3.1in]{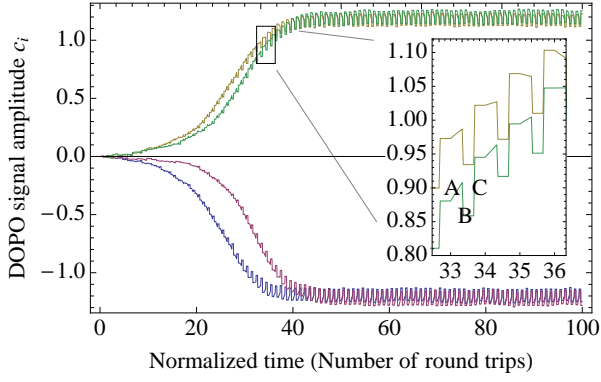}
\caption{\label{fig:maxcut4}Normalized DOPO signal amplitudes as a function of normalized time (in unit of cavity round trip numbers) for a $N=4$ simple MAX-CUT-3 problem. Each color corresponds to the four different DOPOs indexed with $i = 1, \dots, 4$. Small window is enlarged to indicate the status of signal amplitude inside a cavity at three components (as in Fig. \ref{fig:CIM}); A: OPA gain medium (PPLN waveguide), B: out-coupler, and C: injection coupler of the mutual coupling pulse. The two flat regions between B and C and between C and A are the passive propagation in a fiber.}
\end{figure}

\begin{figure}[htbp]
\centering
\includegraphics[width=3.1in]{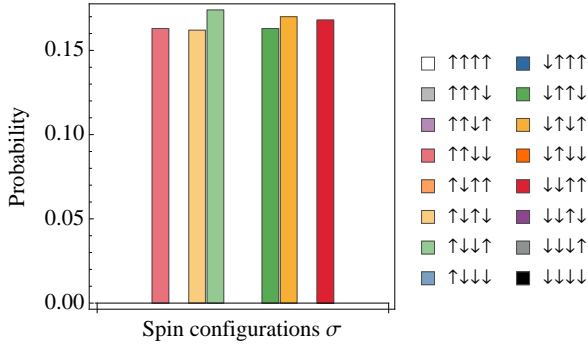}
\caption{\label{fig:4mcut}Distribution of output spin configurations in 1000 trials of numerical simulations against a simple MAX-CUT-3 problem of graph order $N=4$. All trials were successful to find one of the six degenerate ground states.}
\end{figure}

Figure \ref{fig:maxcut4} shows the time evolution of $c_i$ $(i=1,\dots, 4)$ when $p = 1.1$ and $\xi = -0.1$. A correct solution spontaneously emerges after several tens of round trips. The statistics of obtaining different states against 1000 sessions of such a numerical simulation are shown in Fig. \ref{fig:4mcut}, in which six degenerate ground states appear with almost equal probabilities with no errors found.

\subsection{Many-body interaction problem} 
If the interaction is not a standard two-body Ising interaction type but rather a four-body interaction such as 
\begin{eqnarray}\label{eq:4body}
\mathcal{H}=-J_{1234}\sigma_1 \sigma_2 \sigma_3 \sigma_4,
\end{eqnarray}
where $J_{1234}\in \mathbb{R}$, the coupled field into the $i$-th pulse is no longer given by $\sum_j \xi_{ij} \tilde{c}_j$ but by $\xi \tilde{c_j} \tilde{c_k} \tilde{c_l} \ (j, k, l \neq i)$. In this case, the CSDE (\ref{eq:6 coupler 1,2}) can be rewritten to include the four-body coupling term
\begin{eqnarray}\label{eq:4bodyCSDE}
dc_{i}=[(-1+p-c_i^2-s_i^2)c_i+\xi \tilde{c}_j \tilde{c}_k \tilde{c}_l]\, dt \nonumber\\+\frac{1}{A_\mathrm{s}}\sqrt{c_i^2+s_i^2+\frac{1}{2}}dW_i.
\end{eqnarray}

\begin{figure}[htbp]
\centering
\includegraphics[width=3.1in]{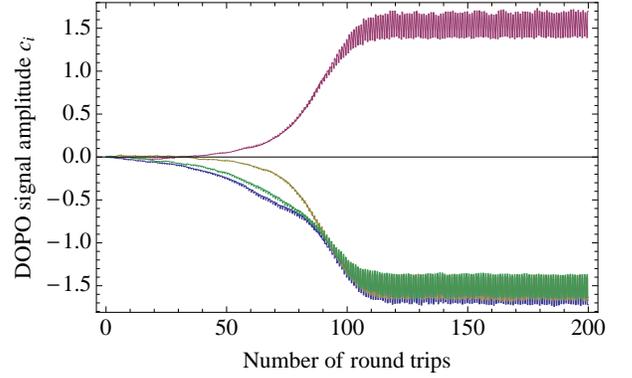}
\caption{\label{fig:amp4}Normalized DOPO pulse amplitudes $c_i$ $(i=1, \dots, 4)$ under the interaction between four-body Ising coupling expressed by Eq. (\ref{eq:4body}).}
\end{figure}

\begin{figure}[htbp]
\centering
\includegraphics[width=3.1in]{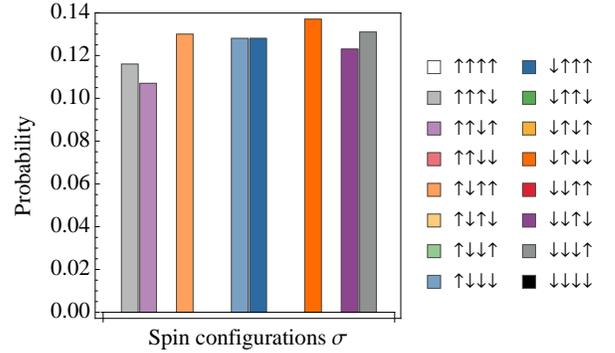}
\caption{\label{fig:4body}Distribution of output spin configurations in 1000 trials of numerical simulation against a four-body Ising model of $N=4$. All trials were successful to find one of the eight degenerate ground states.}
\end{figure}

When the four-body coupling coefficient $J_{1234}$ is $-1$ (multi-body anti-ferromagnetic coupling), there are eight degenerate ground states, i.e., $\left|\uparrow \uparrow \uparrow \downarrow \right\rangle, \left|\uparrow \uparrow \downarrow \uparrow \right\rangle, \left|\uparrow \downarrow \uparrow \uparrow \right\rangle, \left|\downarrow \uparrow \uparrow \uparrow \right\rangle$ and their inverse spin configurations. Figure \ref{fig:amp4} shows the time evolution of $c_i$ $(i=1 ,\dots, 4)$ when $p = 1.1$ and $\xi = -0.1$. One of the eight degenerate ground states emerges spontaneously after several tens of round trips. The statistics of observing different states in 1000 independent sessions of the numerical simulation of Eq. (\ref{eq:4bodyCSDE}) are shown in Fig. \ref{fig:4body}, in which eight degenerate ground states are obtained with no errors found.

\subsection{\label{sec:algorithm}Algorithm description}
In this subsection, we will review the four representative approximation algorithms for MAX-CUT problems.

The Goemans-Williamson algorithm (GW) based on SDP has a $0.87856$-performance guarantee for NP-hard MAX-CUT problems \cite{goemans1995improved}. It achieves the optimal approximation ratio for MAX-CUT problems under the assumptions of $\mathrm{P}\neq \mathrm{NP}$ and the unique games conjecture \cite{khot2007optimal}. The SDP relaxation of the original MAX-CUT problem is a vector-valued optimization problem as maximizing $\ \frac{1}{2}\sum_{i < j}w_{ij}(1 - \vec v_i \cdot \vec v_j), \ \vec v_i \in S^{k-1}$, where $S^{k-1}$ is a unit sphere in $\mathbb{R}^k$ and $k \leq N$ (or $\#V$: number of vertices). There exist polynomial time algorithms to find the optimal solution of this relaxation problem (with error $\varepsilon > 0$), and its value is commonly called the SDP upper bound. A final solution to the original MAX-CUT problem is obtained by projecting the solution vector sets to randomly chosen one-dimensional Euclidean spaces (i.e., dividing the sphere by random hyperplanes).

There are three types of computational complexities of the best-known algorithms for solving the SDP relaxation problem. If a graph with $N$ vertices and $m$ edges is regular, the SDP problem can be approximately solved in almost linear time as $\tilde O (m) = O(m (\log N)^2 \varepsilon^{-4})$ using the matrix multiplicative weights method \cite{arora2007combinatorial}, where $\varepsilon$ represents the accuracy of the obtained solution. However, slower algorithms are required for general graphs. If the edge weights of the graph are all non-negative, the fastest algorithm runs in $\tilde O(Nm) = O(Nm (\log N)^2 \varepsilon^{-3})$ time based on the Lagrangian relaxation method \cite{klein1996efficient}. For graphs with both positive and negative edge weights, the SDP problem is commonly solved using the interior-point method, which scales as $\tilde O(N^{3.5}) = O(N^{3.5} \log(1/\varepsilon))$ \cite{alizadeh1995interior}. Besides, low rank formulation of SDP is effective when the graph is sparse \cite{yamashita2012latest, grippo2012speedp, fujisawa2014petascale}. In our computational experiments, the COPL\_SDP based on the interior point method was used as a solver for MAX-CUT problems \cite{ye1999copl}. The SDP upper bound $U_{\mathrm{SDP}}$ and the solution $C_{\mathrm{GW}}$ were obtained using the following parameters: interior point method was used until the relative gap $r_{\mathrm{gap}} = 1 - P_{\mathrm{obj}}/D_{\mathrm{obj}}$ reached $10^{-3}$, where $P_{\mathrm{obj}}$ and $D_{\mathrm{obj}}$ are the objective functions of the primal and dual of the SDP problem, respectively \cite{benson2000solving}. Random projection onto one-dimensional space was executed $N$ times. 

For many practical applications, heuristic algorithms are more convenient to use, since the GW algorithm generally requires long computation time $\tilde O(N^{3.5})$. Metropolis et al. introduced a simple algorithm that can be used to provide an efficient simulation of a collection of atoms in equilibrium at a given temperature \cite{metropolis1953equation}. Kirkpatrick et al. applied the algorithm to optimization problems by replacing the energy of the atomic system to the cost function of optimization problems and using spin configurations $\sigma$, which is called the simulated annealing algorithm (SA) \cite{kirkpatrick1983optimization}. In each step of this algorithm, a system is given with a random spin flip and the resulting change $\Delta E$ in the energy is computed. If $\Delta E \leq 0$, the spin flip is always accepted, and the configuration with the flipped spin is used as the starting point of the next step. If $\Delta E > 0$, the spin is treated probabilistically, i.e., the probability that the new configuration will be accepted is $P(\Delta E) = \exp(-\Delta E / k_\mathrm{B} T)$ with a control parameter of system temperature $T$. This choice of $P(\Delta E)$ results in the system evolving into an equilibrium Boltzmann distribution. Repeating this procedure, with the temperature $T$ gradually lowered to zero in sufficiently long time, leads spins $\sigma$ to convergence to the lowest energy state. In practical case, with the finite time, the annealing schedule affects the quality of output values. Here in our numerical simulations, the temperature was lowered according to the logarithmic function \cite{hajek1988cooling}. Note that 1 Monte Carlo step corresponds to $N$ trials of spin flip.

Sahni and Gonzalez constructed a greedy algorithm for MAX-CUT problems, which has 1/2-performance guarantee \cite{sahni1976p}, and SG3 is a modified version of it \cite{kahruman2007greedy}. In this algorithm, nodes $V$ are divided into two disjoint subsets $\{S_1, S_2\}$ sequentially. For each iterative process, the node with the maximum score is selected, and it is put into either set $S_1$ or $S_2$ so as to earn larger cuts. Here, the score function of SG3 is defined as $x_i = |\sum_{j \in S_1} w_{ij} - \sum_{j \in S_2} w_{ij}|\ (i = 1, \dots, N)$. It stops when all the edges are evaluated to calculate the score function, thus SG3 scales as $O(m)$.

The power of breakout local search (BLS) appears in the benchmark result for G-set graphs \cite{benlic2013breakout}. It updated almost half of the best solutions in G-set with the specialized data structure for sorting and dedicated procedure to escape from local minima. The algorithm is combination of steepest descent and forced spin flipping: after being trapped by a local minima as a result of steepest descent procedure, three types of forced spin flipping (single, pair, and random) are probabilistically executed according to the vertex influence list (i.e., which vertex will increase the number of cut most when it's flipped) on each subset of partition.

These algorithms are coded in C/C++ and run on a single thread of a single core on a Linux machine with two 6-core Intel Xeon X5650 (2.67 GHz) processors and 94 GB RAM.
The CIM is simulated based on the coupled CSDE (\ref{eq:6 coupler 1,2}) on the same machine.
Note that the computation time of CIM does not mean the simulation time on the Linux machine but corresponds to the actual evolution time of a physical CIM. 

\subsection{\label{sec:SDP}Computational accuracy on G-set instances}
The performance of a CIM with DOPO network was tested on the NP-hard MAX-CUT problems on sparse graphs, so-called G-set \cite{helmberg2000spectral}. These test instances were randomly constructed using a machine-independent graph generator written by G. Rinaldi, with the number of vertices ranging from 800 to 20000, edge density from $0.02\%$ to $6\%$, and topology from random, almost planar, to toroidal. 

The output cut values of running the CIM, SA, GW, and the best known solutions so far we could find \cite{benlic2013breakout, ikuta2015max} for some of G-set graphs are summarized in Table \ref{OPO_SDP}. The results for CIM are obtained in 50 ms, which correspond to the performance of an experimental system after 5000 DOPO cavity round trips. The best result and ensemble average value for 100 trials are shown. Here, the parameters are set to be $p=1.6$, $\xi=-0.06$ and the coupling constant $\xi_{ij} = \xi w_{ij} / \sqrt{\langle k \rangle}$ is normalized by the square root of the graph average degree $\langle k \rangle$. 
The hysteretic optimization method, in which the swinging and decaying Zeeman term that flips the signal amplitude (spin) back and forth, is implemented four times after 10 ms initial free evolution \cite{zarand2002using}. Each hysteretic optimization takes 10 ms so that the total search takes 50 ms.
The result of SA is also obtained in 50 ms for each graph. For GW, the computation time ranged between 2.3 s and $1.1\times 10^5$ s, depending on $N$. The best outputs of the CIM were $1.62 \pm 0.58\ \%$ better than GW but $0.38\pm 0.40\ \%$ worse than SA, and CIM found better cut against GW except for a toroidal graph (g50) and a disconnected random graph (g70).

\begin{table*}
\caption{\label{OPO_SDP}Performance of the coherent Ising machine, simulated annealing and Goemans-Williamson SDP algorithm in solving the MAX-CUT problems on sparse G-set graphs. $\#V$ is the number $N$ of vertices in the graph, $\#E$ is the number $m$ of edges, $U_{\mathrm{SDP}}$ is the optimal solution to the semidefinite relaxation of the MAX-CUT problem when the duality gap reaches to $10^{-3}$,  and $C_{\mathrm{best}}$ is the best known result so far we could find. $C_{\mathrm{GW}}$ is the best solution obtained by $N$ projections after SDP. $C_{\mathrm{SA}}$ and $\langle{C_{\mathrm{SA}}}\rangle$ are the best and average values obtained by SA in 100 trials of 50 ms. $C_{\mathrm{CIM}}$ and $\langle{C_{\mathrm{CIM}}}\rangle$ are the best and average values in CIM in $100$ runs of 50 ms $(=$ 5000 DOPO cavity round trips), respectively. To make comparisons with each other, every cut value $C$ generated from each algorithm is normalized according to $(C+E_{\mathrm{neg}})/(U_{\mathrm{SDP}} + E_{\mathrm{neg}})$, where $E_{\mathrm{neg}}\geq 0$ is the number of negative edges. In the bottom of this table, the average and worst values of all 71 G-set graphs are shown.}
\begin{tabular}{l r r r c c c c c c}
\hline
Graph	&	$\#V$	&	$\#E$	&	$U_{\mathrm{SDP}}$ &	$C_{\mathrm{best}}$	&$C_{\mathrm{GW}}$	&	$C_{\mathrm{SA}}$	&	$\langle{C_{\mathrm{SA}}}\rangle$	&	$C_{\mathrm{CIM}}$	&	$\langle{C_{\mathrm{CIM}}}\rangle$	\\
\hline
g1	&	800	&	19176	&	12083	&	0.9620	&	0.9457	&	0.9620	&	0.9597	&	0.9614	&	0.9570	\\
g6	&	800	&	19176	&	2656	&	0.9607	&	0.9448	&	0.9606	&	0.9592	&	0.9601	&	0.9559	\\
g11	&	800	&	1600	&	629	&	0.9540	&	0.9327	&	0.9526	&	0.9478	&	0.9455	&	0.9370	\\
g14	&	800	&	4694	&	3191	&	0.9602	&	0.9336	&	0.9580	&	0.9544	&	0.9514	&	0.9472	\\
g18	&	800	&	4694	&	1166	&	0.9500	&	0.9282	&	0.9492	&	0.9439	&	0.9434	&	0.9372	\\
g22	&	2000	&	19990	&	14136	&	0.9450	&	0.9191	&	0.9445	&	0.9409	&	0.9405	&	0.9361	\\
g27	&	2000	&	19990	&	4141	&	0.9435	&	0.9174	&	0.9422	&	0.9400	&	0.9390	&	0.9356	\\
g32	&	2000	&	4000	&	1567	&	0.9559	&	0.9272	&	0.9508	&	0.9478	&	0.9424	&	0.9384	\\
g35	&	2000	&	11778	&	8014	&	0.9588	&	0.9292	&	0.9551	&	0.9523	&	0.9471	&	0.9438	\\
g39	&	2000	&	11778	&	2877	&	0.9464	&	0.9226	&	0.9431	&	0.9399	&	0.9364	&	0.9318	\\
g43	&	1000	&	9990	&	7032	&	0.9471	&	0.9292	&	0.9471	&	0.9439	&	0.9458	&	0.9396	\\
g48	&	3000	&	6000	&	6000	&	1.0000	&	1.0000	&	1.0000	&	0.9919	&	1.0000	&	0.9747	\\
g51	&	1000	&	5909	&	4006	&	0.9606	&	0.9333	&	0.9583	&	0.9544	&	0.9506	&	0.9468	\\
g55	&	5000	&	12498	&	11039	&	0.9325	&	0.9006	&	0.9264	&	0.9215	&	0.9193	&	0.9160	\\
g57	&	5000	&	10000	&	3885	&	0.9561	&	0.9237	&	0.9496	&	0.9473	&	0.9419	&	0.9384	\\
g59	&	5000	&	29570	&	7312	&	0.9440	&	0.9148	&	0.9376	&	0.9356	&	0.9308	&	0.9288	\\
g60	&	7000	&	17148	&	15222	&	0.9313	&	0.8989	&	0.9231	&	0.9201	&	0.9191	&	0.9152	\\
g64	&	7000	&	41459	&	10466	&	0.9440	&	0.9143	&	0.9347	&	0.9324	&	0.9320	&	0.9299	\\
g67	&	10000	&	20000	&	7744	&	0.9554	&	0.9215	&	0.9480	&	0.9459	&	0.9411	&	0.9388	\\
g70	&	10000	&	9999	&	9863	&	0.9674	&	0.9633	&	0.9523	&	0.9479	&	0.9515	&	0.9482	\\
g81	&	20000	&	40000	&	15656	&	0.9551	&	0.9195	&	0.9187	&	0.9125	&	0.9393	&	0.9376	\\\hline
Average	&		&		&		&	0.9540	&	0.9303	&	0.9502	&	0.9469	&	0.9464	&	0.9415	\\
Worst	&		&		&		&	0.9313	&	0.8989	&	0.9187	&	0.9125	&	0.9185	&	0.9149	\\
\hline
\end{tabular}
\end{table*}

As the size of optimization problems increases, the average accuracy is important for practical applications. Table \ref{OPO_SDP} shows that for all G-set graphs, the average accuracy in 100 trials is 0.94148 to the SDP upper bound $U_{\mathrm{SDP}}$, i.e., the CIM can find a cut value larger than 0.94148 of the optimal value for the MAX-CUT problems on average, whereas the average accuracy of the GW is 0.93025 and that of the SA is 0.94692. Note that $U_{\mathrm{SDP}}$ is always greater than or equal to the optimal value for each MAX-CUT problem.

\subsection{\label{sec:SA}Computation time on MAX-CUT problems}
In the previous section, the running time of CIM and SA are fixed to estimate the computational accuracy. 
These two algorithms explored the solutions as good as possible in 50 ms. 
Although, if we finish the computation at a certain accuracy, more reasonable computation time can be defined. 
Here, the GW solution was used as the mark of sufficient accuracy because it ensures the 87.856\% of the ground states. 
The CIM and SA then competed the computation time to reach the same values obtained by GW. 
The time and temperature scheduling parameters of the SA were set as follows: Inverse temperature increased with logarithmic function. The number of spin flipping was optimized to be $10^{l}$ times for some $l \in \mathbb{N}$, which requires the minimum computation time to achieve the same accuracy as with the GW.

\begin{figure*}[htbp]
\includegraphics[width=6.1in]{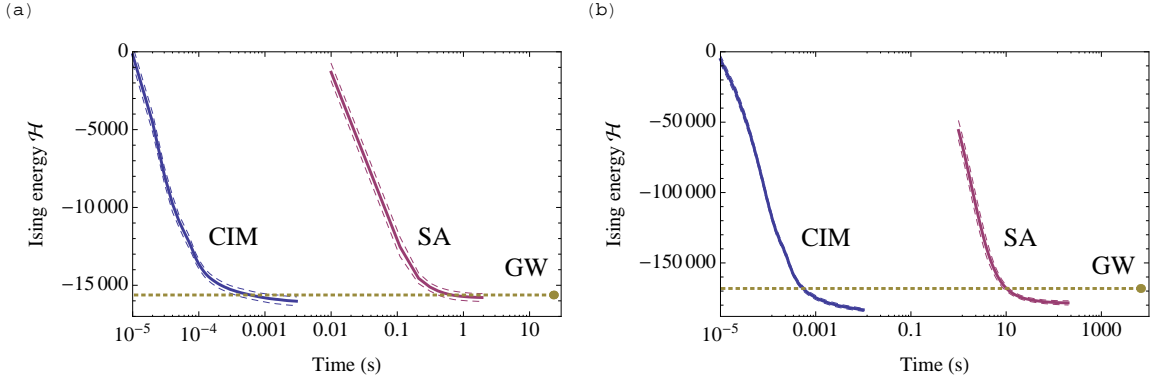}
\caption{\label{fig:energy}Performance comparison of CIM, SA, and GW in solving complete graphs (a) $K_{800}$ and (b) $K_{4000}$, where each edge was randomly weighted $\pm 1$. Each time, bundle of curves depicted average energy (solid line) $\pm$ standard deviations (dashed line) in 100 runs. Dotted line is the accuracy obtained by GW algorithm, whose computation time is shown with dot. The number of spin flip in SA algorithm were $10^5$ for $K_{800}$ and $10^6$ for $K_{4000}$, respectively, to optimize the computation time. 
SA and GW are running on 2.67 GHz Intel Xeon without parallelization and CIM has the cavity circulation frequency of 100 kHz.}
\end{figure*}
Computational experiments were conducted on fully connected complete graphs, denoted by $K_N$, where the number of vertices $N$ ranging from 40 to 20000 and the edges are randomly weighted $\pm 1$. Figure \ref{fig:energy} shows the Ising energy in Eq. (\ref{eq:ising}) as a function of running time. Both CIM and SA run stochastically due to quantum and thermal noise, respectively, the ensemble average of energies are calculated as follows: For the CIM, the energy of all 100 runs was averaged at each round trip. For the SA, the averaged energy was calculated at each point on the time axis with an interpolated value from real time sampling. The parameters for CIM are chosen to be $p=0.2$, $\xi=-0.03$ (for $N=800$), and $\xi=-0.003$ (for $N=4000$). In Fig. \ref{fig:energy} (a), where $N = 800$, the GW achieved an energy equal to $-15624$ in $22.96\ \mbox{s}$, while the CIM and SA reached the same energy in $6.1 \times 10^{-4}\ \mbox{s}$ and in $0.671\ \mbox{s}$. 
In Fig. \ref{fig:energy} (b), the GW achieved an energy of $-168160$ in $6646.25\ \mbox{s}$, while the CIM reached the same energy in $5.5 \times 10^{-4}\ \mbox{s}$ and the SA did so in $10.1\ \mbox{s}$. 

Note that this result of SA and GW comes from a specific computer configuration as mentioned in Sec. \ref{sec:algorithm}. There is room for an improvement in the computation time in constant factor due to cases like using faster CPUs or parallelized codes. 
Similarly, the computation time of CIM also depends on the system configuration and can be made faster when we use the higher clock frequency. 
In this sense, the ratios between time of CIM and that of the other algorithms are arbitrary. 
Thus we should study the computation time scaling as a function of the problem size. 

\begin{figure*}[htbp]
\centering
\includegraphics[width=6.1in]{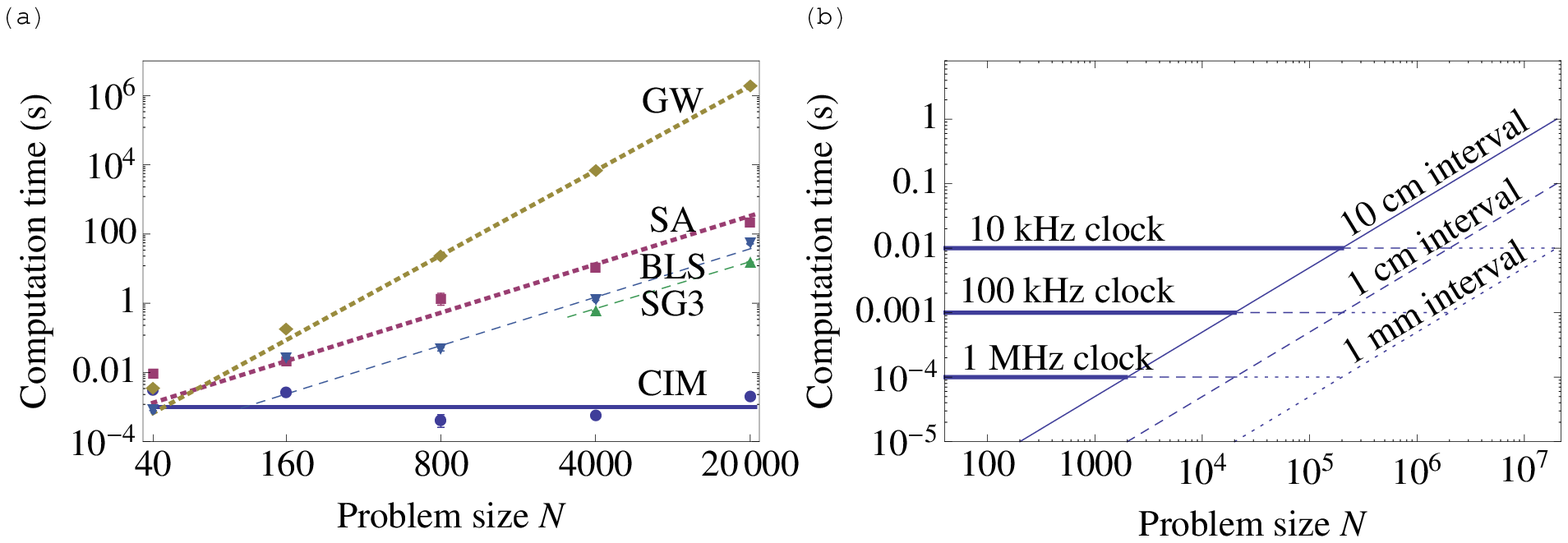}
\caption{\label{fig:time_scaling}(a) Computation time of coherent Ising machine (CIM) empirically scales as $O(1)$, while that of simulated annealing (SA), and Goemans-Williamson SDP (GW) fitted well to lines indicating $O(N^2)$ and $O(N^{3.5})$, respectively. Computation time of SG3 and BLS also fit to $O(N^2)$ with a difference of constant factor. The computation time of CIM, SA, SG3, and BLS is defined as the time to reach the same accuracy achieved by GW (without I/O time). Data points of CIM, SA, SG3, and BLS were calculated by averaging 100 runs. Only for $N = 800$ case, 100 randomly $\pm 1$ weighted complete graphs are generated. The results for $N = 800$ show the average computation time for 100 graphs with error bars indicating standard deviations (which are distributed as in Fig. \ref{fig:tod} (b)). 
Note that the computation time of CIM is evaluated by (number of round trips) $\times$ (10 $\mu$s) as in Sec. \ref{sec:spec}. (b) The computation time of CIM when the DOPO cavity circulation frequency \{10 kHz, 100 kHz, 1 MHz\} and pulse interval \{10 cm (solid line), 1 cm (dashed line), 1 mm (dotted line)\} are changed. Computation time of CIM scales as $O(N)$ if we vary the fiber length proportional to $N$ with fixed pulse interval.
}
\end{figure*}
Figure \ref{fig:time_scaling} (a) shows the computation time versus problem size (number of vertices). The computation time is defined as the CPU time to solve a given MAX-CUT problem in complete graph for GW; as the CPU time to reach the same accuracy as GW for SA, SG3, and BLS; and as the time estimated by the (number of round trips) $\times$ (cavity round trip time) to obtain the same accuracy as GW for CIM. The preparation time needed to input $J_{ij}$ into the computing system, i.e., the graph I/O time, is not included. 
For complete graphs of $N \leq 20000$, the CIM exhibits a problem-size independent computation time of less than $10^{-3}\ \mathrm{s}$ if we assume the fixed cavity circulation frequency of 100 kHz and pulse interval of 10 cm. 
This means the target accuracy is obtained in the constant number of round trips. It indicate that the computation time of CIM is determined by the turn-on delay time of the DOPO network oscillation, which in turn depends on the round trip time and the pump rate. 

In Figure \ref{fig:time_scaling} (b), the computation time of the CIM with different system clock frequency 
and pulse spacing are shown (see the Sec. \ref{sec:spec} for the definition). 
Since the solutions are obtained in a constant number of cavity round trips, the computation time is pulse spacing independent but linearly depends on the clock frequency, i.e., cavity circulation frequency. 
The number of pulses accommodated in the fiber can be changed to vary the pulse spacing under the fixed clock frequency. 
On the other hand, when the pulse spacing is fixed and the fiber length is varied, the maximum number of pulses should be increased in proportional to the fiber length. 

\begin{figure*}
\centering
\includegraphics[width=6.1in]{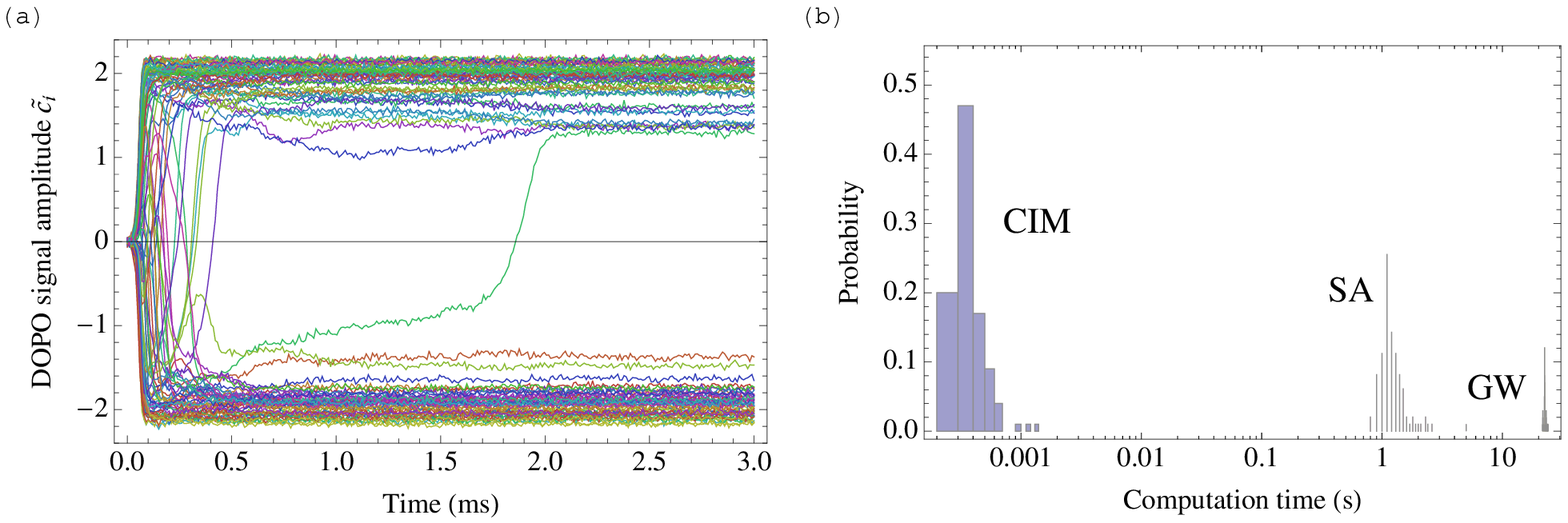}
\caption{\label{fig:tod}Detailed data for $N = 800$ randomly weighted complete graphs (appeared in Fig. \ref{fig:energy} (a)). (a) Time evolution of DOPO amplitudes of the graph of $N = 800$. 
Only 100 pulses out of $N = 800$ signal pulses are shown. (b)Histogram of computation times for solving 100 randomly weighted complete graphs of size $N=800$ by the CIM, SA, and GW.}
\end{figure*}

Going back to the Figure \ref{fig:time_scaling} (a), the time complexity $O(N^{3.5})$ for the GW is dominated by the interior-point method in the Goemans-Williamson algorithm. The SA seems to scale in $O(N^{2})$, which indicates that it requires the number of spin flips to be proportional to $N$ (i.e., constant Monte Carlo steps) to achieve the optimal performance. Each spin flip costs a computation time proportional to the degree $k_i$, where $k_i$ is equal to $N-1$ for all $i \in V$ in the case of complete graphs. Thus, the computation time scales as $O(N \langle k \rangle) = O(N^2)$ for the SA in the complete graphs. Note that CIM and SA didn't always reach the energy obtained by GW for the graph of $N = 40$, half of the 100 runs of stochastic algorithms were post-selected to reach that value. SG3 scales as $O(m) = O(N^2)$ in Fig. \ref{fig:time_scaling} (a), but the values for $N = 40, 160, 800$ are not shown because it didn't reach the accuracy reached by the GW solution. BLS exhibits competitive performance against SA. 
Besides, the DOPO amplitudes in CIM evolve as in Fig. \ref{fig:tod} (a) when $N = 800$. The distribution of computation time for 100 randomly weighted complete graphs of $N = 800$ is also shown in Fig. \ref{fig:tod} (b). 

\begin{figure}[htbp]
\centering
\includegraphics[width=3.1in]{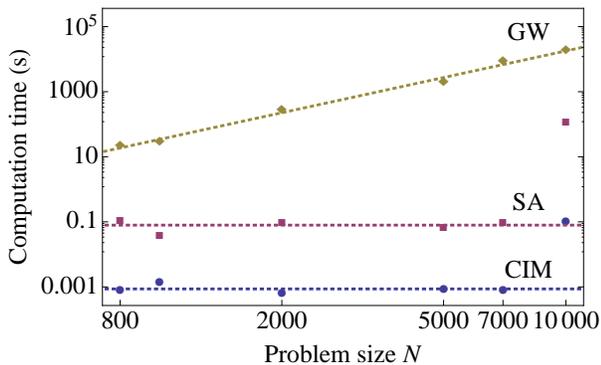}
\caption{\label{fig:gset}Computation time of coherent Ising machine, simulated annealing algorithm, and Goemans-Williamson SDP algorithm on random graphs in G-set instances. The computation time for CIM and SA is defined as the time to reach the same accuracy achieved by GW (without I/O time). 
Note that the computation time of CIM is evaluated by (number of round trips) $\times$ (10 $\mu$s) as in Sec. \ref{sec:spec}.}
\end{figure}
Computation time for the random graphs in G-set instances is also studied. Here the subset of graphs in which MAX-CUT problems can be solved in polynomial time (i.e., planar graphs, weakly bipartite graphs, positive weighted graphs without a long odd cycle, and graphs with integer edge weight bounded by $N$ and fixed genus) are excluded. 
The execution time of CIM is evaluated under the machine spec described in Sec. \ref{sec:spec} with $p = 0.2$, $\xi = -0.06$, and $\xi_{ij} = \xi w_{ij} / \sqrt{\langle k \rangle}$. Again, the computation time of SA and CIM is the actual time (without graph file I/O) to obtain the same accuracy of solution as GW. 
Figure \ref{fig:gset} shows the computation time as functions of the problem size $N$. The computational cost of interior point method dominates the GW algorithm. (Note that G-set contains graphs with both positive and negative edge weights so that we must use the slowest interior point method.) 
Then the computation time is almost constant for both SA and CIM. The computation time of SA with constant Monte Carlo step is expected to scale $O(N \langle k \rangle) \sim O(1)$ (here for the random graphs in G-set, $\langle k \rangle \sim O(N^{-1.09}$)). The computation time of CIM here is governed by a turn-on delay time of the DOPO network to reach a steady state oscillation condition, which is constant for varying values of $N$ as mentioned above \cite{takata2012transient}.

\section{\label{sec:outro}Summary and discussion}
The potential for solving NP-hard problems using a CIM was numerically studied by conducting computational experiments using the MAX-CUT problems on sparse G-set graphs and fully connected complete graphs of order up to $2 \times 10^4$. With the normalized pump rate and coupling coefficient $p=0.2$ and $\xi=-0.06$, the CIM achieved a good approximation rate of 0.94148 on average and found better cut compared to the GW for 69 out of 71 graphs in G-set. The computation time for this sparse graph set, including few sessions of hysteretic optimization, is estimated as $50$ ms.  
The time scaling was also tested on complete graphs of number of vertices up to $2 \times 10^4$ and number of edges up to $10^8$. The results imply that CIM achieves empirically constant time scaling in a fixed system clock frequency, i.e., the fixed cavity circulation frequency (fiber length),  
while SA, SG3, and BLS scale as $O(N^2)$ and GW scales as $O(N^{3.5})$. 
Those results suggest that CIM may find applications in high-speed computation for various combinatorial optimization problems, in particular for temporal networks.

The present simulation results do not mean that the CIM can get a reasonably accurate solution by a constant time for arbitrary large problem size. As mentioned already, in the CIM based on a fiber ring resonator, the number of DOPO pulses is determined by the the fiber length and the pulse spacing. In order to implement $2 \times 10^5$ and $2 \times 10^6$ DOPO pulses in the $20\ \mathrm{km}$ fiber ring cavity, we must use a pulse repetition frequency to 2 GHz and 20 GHz, respectively. This is a challenge for both optical components and electronic components of CIM, but certainly within a reach in current technologies. 

\begin{acknowledgments}
The authors would like to thank K. Inoue, H. Takesue, K. Aihara, A. Marandi, P. McMahon, T. Leleu, S. Tamate, K. Yan, Z. Wang, and K. Takata for their useful discussions. 
This project is supported by the ImPACT program of the Japanese Cabinet Office.
\end{acknowledgments}

\end{document}